\documentclass{article}
\usepackage{graphicx}

\newcommand{\graphHeight}{5cm}
\newcommand{\heurHeight}{4.4cm}
\newcommand{\graphDraw}{1}

\newcommand{\wH}[2]{{h_{#2\mid #1}}}
\newcommand{\wP}[2]{p_{#2\mid #1}}
\newcommand{\wPs}[0]{P_{\mathrm{s}}}
\newcommand{\wPv}[0]{P_{\mathrm{v}}}
\newcommand{\qgcc}[1]{{q_{#1}^{V}}}

\newcommand{\degcc}[1]{{w_{#1}^{V}}}

\newcommand{\GCC}[1]{{\mathrm{GCC}_{#1}}}
\newcommand{\GOUT}[1]{{\mathrm{GOUT}_{#1}}}
\newcommand{\GIN}[1]{{\mathrm{GIN}_{#1}}}
\newcommand{\GWCC}[1]{{\mathrm{GWCC}_{#1}}}
\newcommand{\GSCC}[1]{{\mathrm{GSCC}_{#1}}}
\newcommand{\gcc}[1]{\theta_{#1}}
\newcommand{\gin}[1]{{\theta_{#1}^{\mathrm{in}}}}
\newcommand{\gout}[1]{{\theta_{#1}^{\mathrm{out}}}}
\newcommand{\qin}[1]{{q_{#1}^{\mathrm{in}}}}
\newcommand{\qout}[1]{{q_{#1}^{\mathrm{out}}}}

\newcommand{\dein}[1]{{w_{#1}^{\mathrm{in}}}}
\newcommand{\deout}[1]{{w_{#1}^{\mathrm{out}}}}

\begin{document}

\title{A Dissemination Strategy for Immunizing Scale-Free Networks}

\author{Alexandre~O.~Stauffer\\
Valmir~C.~Barbosa\thanks{Corresponding author (valmir@cos.ufrj.br).}\\
\\
Universidade Federal do Rio de Janeiro\\
Programa de Engenharia de Sistemas e Computa\c c\~ao, COPPE\\
Caixa Postal 68511\\
21941-972 Rio de Janeiro - RJ, Brazil}

\date{}

\maketitle

\begin{abstract}
We consider the problem of distributing a vaccine for immunizing a scale-free
network against a given virus or worm. We introduce a new method, based on
vaccine dissemination, that seems to reflect more accurately what is expected to
occur in real-world networks. Also, since the dissemination is performed using
only local information, the method can be easily employed in practice. Using a
random-graph framework, we analyze our method both mathematically and by means
of simulations. We demonstrate its efficacy regarding the trade-off between the
expected number of nodes that receive the vaccine and the network's resulting
vulnerability to develop an epidemic as the virus or worm attempts to infect one
of its nodes. For some scenarios, the new method is seen to render the network
practically invulnerable to attacks while requiring only a small fraction of the
nodes to receive the vaccine.

\end{abstract}

\section{Introduction} \label{sec:intro}

The term ``scale-free'' is widely used to designate the class of networks that
have node degrees distributed as a power law \cite{barabasi1999,newman2003b},
according to which the probability that a randomly chosen node has degree $a$ is
proportional to $a^{-\tau}$ for some parameter $\tau>0$. There has been a recent
surge of interest in scale-free networks, as a great variety of real-world
networks, like the Internet, the WWW, social networks, and
scientific-collaboration networks, have been empirically observed to have
node-degree distributions that approximately follow a power law
\cite{faloutsos1999,albert2002}. In contrast with the classical random-graph
model introduced by Erd\H{o}s and Rényi, whose node-degree distribution is the
Poisson distribution, therefore sharply concentrated around its mean value
\cite{erdos1959,bollobas2002}, scale-free networks normally contain nodes with
a wide range of degrees, typically with a few nodes of extremely high degrees
coexisting with a plethora of low-degree nodes.

In this paper we consider the problem of preventing viruses or worms from
spreading on scale-free computer networks. The fact that node degrees are in
this case distributed according to a power law has profound impact on the way
the network operates. In particular, it makes the problem of fighting the
proliferation of viruses and other infections much more challenging, since the
presence of high-degree nodes dramatically increases the rate at which a virus
may propagate \cite{satorras2001,satorras2003}. For this reason, instead of
combating the proliferation of a virus in an already infected network, we
consider a preventive immunization strategy, which consists of distributing the
appropriate vaccine to a small subset of the network's nodes, striving to
immunize those nodes that can more efficiently block the spread of a future
infection. The goal of this approach is to distribute the vaccine to as few
nodes as possible while making the network invulnerable to an epidemic, that is,
to the occurrence of a state in which a relatively large number of nodes is
infected.

We can measure the efficacy of an immunization strategy by two indicators: the
expected spread, which is the expected fraction of the network's nodes that
receive the vaccine, and the expected vulnerability, which is the expected
fraction of the network's nodes that may become infected when the virus attempts
to infect a randomly chosen node of the immunized network. Clearly, these two
indicators are strongly influenced by how we select the nodes to receive the
vaccine. A simple rule for choosing these nodes is to randomly select a given
fraction of the network's nodes \cite{albert2000,cohen2000,satorras2003}. When
applied to scale-free networks, we know that this rule normally gives
unsatisfactory results, as it only achieves a reasonably small expected
vulnerability for prohibitively high expected spreads. An alternative rule
consists of distributing the vaccine to all the nodes that have degrees greater
than a given value \cite{albert2000,cohen2001,satorras2003}. Despite being more
efficient for scale-free networks than the previous strategy, as it achieves
quite a small expected vulnerability with only a modest expected spread,
applying this rule to real-world networks is known to be usually difficult
\cite{cohen2003}. The use of this rule demands global knowledge regarding the
location of the nodes having the highest degrees, while the nodes of many
real-world networks may only be assumed to have information that can be directly
inferred from their immediate neighborhoods. Yet another alternative is to
randomly choose some of the network's nodes and, for each of them, to immunize a
randomly chosen fraction of its neighbors \cite{cohen2003}. This rule, however,
and in fact the previous two as well, seems hard to implement in practice on
computer networks, since apparently it requires that the vaccine be somehow
transmitted to a given fraction of the network's nodes by means other than the
network's own.

In this paper, we assume that the vaccine enters the network at a single node,
called the originator. We attribute to this node the responsibility of starting
the dissemination of the vaccine by initiating the method called heuristic
flooding for disseminating information in networks \cite{stauffer2004}. Let $u$
be the originator. For each neighbor $v$ of $u$, this method prescribes that $u$
forward the vaccine to $v$ with probability given by a heuristic function
$h(a,b)$, where $a$ and $b$ are, respectively, the degrees of $u$ and
$v$.\footnote{We assume $h(a,b)=0$ if $a=0$ or $b=0$.} Each of the nodes that
receive the vaccine, when receiving it for the first time, proceeds likewise and
probabilistically forwards the vaccine to its own neighbors. By not requiring
that the nodes of the network have information beyond what can be inferred from
their immediate neighborhoods, this strategy can be easily used in practice.
Furthermore, it represents more accurately what occurs in real scenarios, since
it does not rely on the prior selection of nodes that characterizes all the
three immunization strategies mentioned above, but rather assumes that the
vaccine spreads out of a single node (say, the very site of its development or
the site responsible for its distribution) via a heuristically controlled form
of flooding. With this set of characteristics that, in essence, make it
independent of any network-wide properties, the new strategy is to our knowledge
the first of a kind.
 
We organize the remainder of the paper as follows. In Section~\ref{sec:math}, we
use a random-graph framework and the formalism introduced in
\cite{molloy1995,molloy1998,newman2001}, whose details are discussed as they are
needed, to obtain mathematical results for the aforementioned efficacy
indicators. We utilize our analytical results in Section~\ref{sec:heur} to
discover the properties that an ideal heuristic function should have to be
efficient. We then introduce a heuristic function that seeks to approximate this
ideal and therefore can be used to disseminate the vaccine. In
Section~\ref{sec:sim} we discuss simulation results on random graphs having node
degrees distributed according to a power law. Our results reveal that this
heuristic function performs very attractively for the ranges of $\tau$ (the
distribution's parameter) that typically are thought to hold for networks like
the Internet. They also agree satisfactorily with our analytical predictions. We
conclude in Section~\ref{sec:conc}.

\section{Mathematical analysis} \label{sec:math}

Let $G$ be a random graph having $n$ nodes, whose degrees are distributed
independently from one another and identically to a random variable $K_G$. We
assume that the nodes of $G$ are interconnected in an independent way given
their degrees, which therefore remain independent. We base our mathematical
analysis of this section on the formalism introduced in \cite{newman2001} and
target the case in which $G$ has a formally infinite number of nodes.

Let $P_G(a)$ be the probability that a randomly chosen node of $G$ has degree
$a$, i.e., the probability that $K_G=a$. The average degree in $G$, denoted by
$Z_G$, is clearly
\begin{equation}
   Z_G = \sum_{a=0}^{n-1} a P_G(a).
\end{equation}
Given that the degrees of two adjacent nodes are independent from each other,
the probability that some node's neighbor has degree $b$ is identical to the
expected fraction of edges incident to degree-$b$ nodes, which is given by
\begin{equation}
   \frac{b P_G(b)}{\sum_{a=0}^{n-1}a P_G(a)} = \frac{b P_G(b)}{Z_G}.
   \label{eq:neigh}
\end{equation}

From \cite{molloy1995,newman2001}, a necessary and sufficient condition for a
size-$\Theta(n)$ connected component to almost surely exist in $G$ is that
\begin{equation}
   \sum_{b=1}^{n-1} (b-1) \frac{b P_G(b)}{Z_G} > 1,
   \label{eq:phase}
\end{equation}
which intuitively means that, given a randomly chosen node $u$ of $G$, a
size-$\Theta(n)$ connected component exists almost surely if and only if a
neighbor of $u$ is expected to have more than $1$ neighbor besides $u$. When
(\ref{eq:phase}) is satisfied, we denote the size-$\Theta(n)$ connected
component of $G$ (its giant connected component) by $\GCC{G}$. Also, all the
other connected components of $G$ are small with high probability, comprising
only $o(n)$ nodes, and $G$ is said to be above the phase transition that gives
rise to $\GCC{G}$. On the other hand, when (\ref{eq:phase}) is not satisfied,
$G$ is said to be below the phase transition that gives rise to $\GCC{G}$ and
all of its connected components are small with high probability, consisting each
of $o(n)$ nodes.

Given a randomly chosen node $u$ of $G$ and a neighbor $v$ of $u$, we define the
reach of $u$ through $v$ as the set of nodes that can be reached by a path
starting at $u$ and whose first edge is $(u,v)$. A node belongs to $\GCC{G}$ if
and only if it has at least one neighbor through which its reach contains a
large, size-$\Theta(n)$ number of nodes. Let $q$ be the probability that a node
has a small, size-$o(n)$ reach through a given neighbor. The probability that a
degree-$a$ node belongs to $\GCC{G}$ is then $1-q^a$, and the probability that a
randomly chosen node of $G$ belongs to $\GCC{G}$, which we denote by $\gcc{G}$,
is
\begin{equation}
   \gcc{G} = 1 - \sum_{a=0}^{n-1} q^a P_G(a).
   \label{eq:gccG}
\end{equation}
The probability $q$ that $u$ has a small, size-$o(n)$ reach through $v$ can be
obtained from the probability that $v$ itself has a small, size-$o(n)$ reach
through each of its other neighbors (i.e., excluding $u$). Since the probability
that two neighbors of $u$ have another common neighbor (i.e., besides $u$)
varies with $n$ proportionally to $n^{-1}$ \cite{newman2001}, which for large
$n$ is negligible, the probability that $v$ has a small, size-$o(n)$ reach
through a given neighbor is also $q$, thus leading to
\begin{equation}
   q = \sum_{b=1}^{n-1} q^{b-1} \frac{b P_G(b)}{Z_G}.
\end{equation}
This equation can be solved numerically and then used in (\ref{eq:gccG}) to
obtain $\gcc{G}$.

From now on, we assume that $G$ is above the phase transition and, therefore,
$\GCC{G}$ exists with high probability. Furthermore, since $G$ can be
unconnected and real-world computer networks are normally connected, we assume
that it is the graph induced by $\GCC{G}$, rather than $G$ itself, that models
the network, and also condition the remainder of our analysis accordingly.

\subsection{Expected spread} \label{sec:spread}

In this section, we calculate the expected spread in $\GCC{G}$, which is denoted
by $\wPs$ and consists of the expected fraction of the nodes of $\GCC{G}$ that
are immunized when a vaccine is distributed using the heuristic flooding
described in Section~\ref{sec:intro}. We resort to the same method of analysis
developed in \cite{stauffer2004}. Let $S$ be a directed subgraph of $G$ that
spans all the nodes of $G$. For a degree-$a$ node $u$ and a degree-$b$ neighbor
$v$ of $u$ in $G$, the probability that the directed edge $(u \to v)$ exists in
$S$ is given by $h(a,b)$, the heuristic function employed during the vaccine
dissemination. Before proceeding to the calculation of $\wPs$, we pause for a
brief study of $S$.

The neighbors of a node $u$ in $S$ can be classified into two different types:
the in-neighbors, those from which an edge exists directed toward $u$; and the
out-neighbors, those toward which an edge exists directed from $u$. If a
directed path exists starting at some node $u$ and ending at another node $v$,
then we say that $u$ reaches $v$ in $S$ or that $v$ is in the reach of $u$ in
$S$. Note that, if $u$ receives the vaccine, then the reach of  $u$ in $S$ is
part of the set of nodes that become immunized.

The connected components of a directed graph can also be of two basic types.
First, there are the weakly  connected components, which are constituted by the
nodes that can reach one another by undirected paths, i.e., paths for which the
directions of the edges are disregarded. The other type is that of the strongly
connected components, each comprising a maximal set of nodes that can both reach
and be reached from one another.

Similarly to the case of the undirected graph $G$, there is a criterion for
deciding whether $S$ almost surely has a size-$\Theta(n)$ weakly connected
component, commonly known as the giant weakly connected component of $S$,
denoted by $\GWCC{S}$. Likewise, there is another criterion according to which
$S$ almost surely has a size-$\Theta(n)$ strongly connected component, commonly
referred to as the giant strongly connected component, denoted by $\GSCC{S}$.
Clearly, when both $\GWCC{S}$ and $\GSCC{S}$ exist, as we henceforth assume, all
the nodes of $\GSCC{S}$ belong also to $\GWCC{S}$, and all the nodes of
$\GWCC{S}$ belong also to $\GCC{G}$.

\begin{figure*}[!t]
   \centering
   \begin{tabular}{c}
   \includegraphics[scale=\graphDraw]{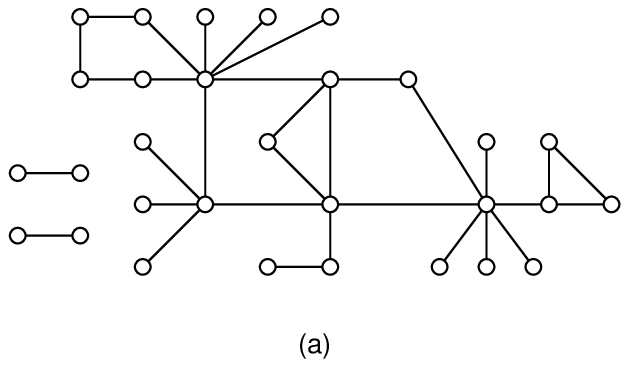}\\
   \\
   \includegraphics[scale=\graphDraw]{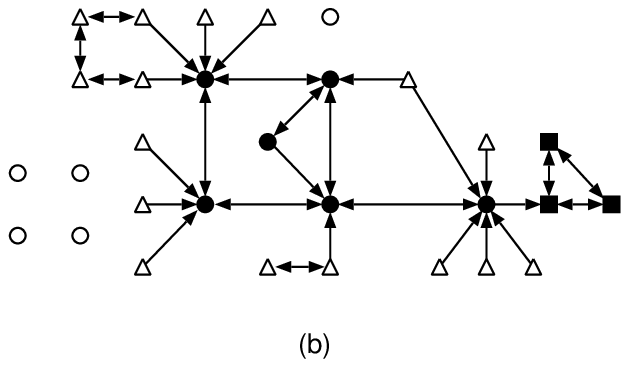}
   \end{tabular}
   \caption{A $G$ instance having a power-law node-degree distribution with
$\tau=2.1$ (a) and one possible instance of the directed subgraph $S$ of the $G$
instance (b). Part~(b) also shows the nodes belonging to $\GSCC{S}$ (filled
circles), $\GIN{S}$ (filled circles and triangles), and $\GOUT{S}$ (filled
circles and filled squares).}
   \label{fig:ggi}
\end{figure*}

Since $\GSCC{S}$ exists by assumption, we can define two other size-$\Theta(n)$
connected components of $S$, which we refer to as the giant in-component
($\GIN{S}$), formed by the nodes that can reach $\GSCC{S}$, and the giant
out-component ($\GOUT{S}$), formed by the nodes reachable from $\GSCC{S}$. Note
that, by definition, the nodes of $\GSCC{S}$ belong also to both $\GIN{S}$ and
$\GOUT{S}$. We denote by $\gin{S}$ and $\gout{S}$ the expected fraction of the
nodes of $G$ that belong to, respectively, $\GIN{S}$ and $\GOUT{S}$.
Figure~\ref{fig:ggi} illustrates an instance of graph $G$ having a power-law
node-degree distribution with $\tau=2.1$ (part~(a)) and a possible instance of
its directed subgraph $S$ (part~(b)).

Assuming that the originator is randomly chosen among the nodes of $\GCC{G}$,
the vaccine is guaranteed to be distributed to a size-$\Theta(n)$ set of nodes
if the originator belongs to $\GIN{S}$, which happens with probability
$\gin{S}/\gcc{G}$. When this is the case, the nodes that receive the vaccine
either belong to $\GOUT{S}$, corresponding to a fraction $\gout{S}/\gcc{G}$ of
the nodes of $\GCC{G}$, or are not in $\GOUT{S}$ despite being reachable from
the originator, and then amount to a small, size-$o(n)$ number of nodes.
Neglecting the latter nodes is equivalent to assuming that nodes receive the
vaccine only if the originator is in $\GIN{S}$. In this case, only the nodes in
$\GOUT{S}$ receive the vaccine and we have
\begin{equation}
   \wPs = \frac{\gin{S}\gout{S}}{\gcc{G}^2}.
   \label{eq:wPi}
\end{equation}

In order to obtain $\gin{S}$, recall that the nodes of $\GIN{S}$ are the only
ones that have a non-negligible reach. Considering a degree-$a$ node $u$ of $G$
and a degree-$b$ neighbor $v$ of $u$ in $G$, we say that $v$ is a dead end with
respect to $u$ in $S$ if either $(u \to v)$ is not an edge of $S$, or it is but
the reach of $u$ through $v$ in $S$ is negligible, consisting of only $o(n)$
nodes. Denoting by $\qin{b}$ the conditional probability that the reach of $u$
through $v$ in $S$ is negligible given that $u$ is an in-neighbor of $v$ in $S$,
 we obtain the probability that $v$ is a dead end with respect to $u$ in $S$,
which is
\begin{equation}
   1-h(a,b)+h(a,b)\qin{b}.
\end{equation}
And since the probability that $v$ has degree $b$ is given by (\ref{eq:neigh}),
the probability that a given neighbor of a degree-$a$ node is a dead end with
respect to it in $S$, which we denote by $\dein{a}$, is 
\begin{equation}
   \dein{a} = \sum_{b=1}^{n-1} \left(1-h(a,b)+h(a,b)\qin{b}\right) \frac{bP_G(b)}{Z_G}.
   \label{eq:dein}
\end{equation}
Because a node belongs to $\GIN{S}$ if and only if at least one of its neighbors
in $G$ is not a dead end with respect to it in $S$, we arrive at
\begin{equation}
   \gin{S} = 1 - \sum_{a=0}^{n-1} (\dein{a})^a P_G(a).
   \label{eq:gin}
\end{equation}

As a means to calculate $\qin{b}$, let us consider a degree-$b$ node $v$ of $G$
reached by following a directed edge $(u \to v)$ of $S$. The reach of $u$
through $v$ in $S$ is negligible, which happens with probability $\qin{b}$, if
and only if all of the other $b-1$ neighbors of $v$ in $G$ (i.e., excluding $u$)
are themselves dead ends with respect to $v$ in $S$. This clearly leads to
\begin{equation}
   \qin{b} = (\dein{b})^{b-1}.
   \label{eq:qin}
\end{equation}
Equations (\ref{eq:dein}) and (\ref{eq:qin}) can be put together to yield
another equation where $\dein{a}$ is a function of all the other $\dein{}$'s.
This equation can then be solved numerically to obtain $\gin{S}$ via
(\ref{eq:gin}).

We can follow a completely analogous derivation and obtain $\gout{S}$ by noting
that a node belongs to $\GOUT{S}$ if and only if it can be reached from a
size-$\Theta(n)$ set of nodes. Let $u$ be a degree-$a$ node of $G$ and $v$ a
neighbor of $u$ in $G$. We denote by $\deout{a}$ the probability that either
$u$ is not an out-neighbor of $v$ in $S$ or is but the number of nodes that can
reach $u$ through $v$ in $S$ is small, consisting of only $o(n)$ nodes. Also, we
denote by $\qout{b}$ the conditional probability that the number of nodes that
can reach $u$ through $v$ in $S$ is small, given that the degree of $v$ in $G$
is $b$ and $u$ is an out-neighbor of $v$. In a way analogous to the one that led
to (\ref{eq:dein}), (\ref{eq:gin}), and (\ref{eq:qin}), we obtain
\begin{equation}
   \deout{a} = \sum_{b=1}^{n-1} \left(1-h(b,a)+h(b,a)\qout{b}\right)
   \frac{bP_G(b)}{Z_G}, \label{eq:deout}
\end{equation}
\begin{equation}
   \gout{S} = 1 - \sum_{a=0}^{n-1} (\deout{a})^a P_G(a),
   \label{eq:gout}
\end{equation}
and
\begin{equation}
   \qout{b} = (\deout{b})^{b-1}.
   \label{eq:qout}
\end{equation}
Also, and identically to the derivation of $\gin{S}$, we can unify
(\ref{eq:deout}) and (\ref{eq:qout}) and calculate the value of each
$\deout{a}$ numerically to obtain $\gout{S}$ via (\ref{eq:gout}).

\subsection{Expected vulnerability} \label{sec:vulnerability}

Consistently with the simplifying assumptions of Section~\ref{sec:spread}, we
keep assuming that no node is immunized when the originator does not belong to
$\GIN{S}$. When this happens, all nodes of $\GCC{G}$ remain vulnerable to the
virus, and if the virus infects a node of $\GCC{G}$ it may propagate until the
entire $\GCC{G}$ is infected. Let us analyze the case in which the originator
does belong to $\GIN{S}$.

\begin{figure*}[!t]
   \centering
   \includegraphics[scale=\graphDraw]{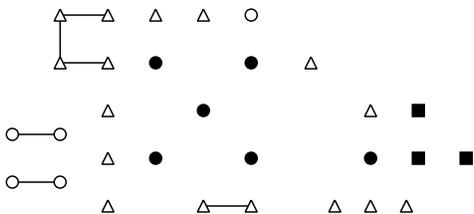}
   \caption{The graph $V$ that corresponds to the $G$ and $S$ instances of
Figure~\ref{fig:ggi}. Nodes represented by filled circles or filled squares
belong to $\GOUT{S}$.}
   \label{fig:ggc}
\end{figure*}

As before, we assume that only the nodes of $\GOUT{S}$ receive the vaccine. Let
$V$ be an undirected subgraph of $G$ that spans all the nodes of $G$, and let an
edge $(u,v)$ of $G$ belong to $V$ if and only if neither $u$ nor $v$ belongs to
$\GOUT{S}$. That is, given a certain instance of the subgraph $S$, subgraph $V$
contains all the edges of $G$ that are not incident to nodes of $\GOUT{S}$.
Clearly, the edges of $V$ represent the edges through which the virus may
propagate if it reaches either of an edge's (unimmunized) end nodes.
Figure~\ref{fig:ggc} illustrates the subgraph $V$ corresponding to the $G$ and
$S$ instances of Figure~\ref{fig:ggi}.

Once again, and similarly to the case of $G$, a criterion exists for deciding
whether a size-$\Theta(n)$ connected component almost surely exists in $V$. We
denote such a component by $\GCC{V}$. When it does exist, and since all the
other connected components of $V$ contain with high probability only $o(n)$
nodes (which we again neglect), a virus may only proliferate into a large,
size-$\Theta(n)$ set of nodes if it first infects a node of $\GCC{V}$. This, of
course, is predicated upon the originator being in $\GIN{S}$ and dissemination
taking place exclusively inside $\GOUT{S}$, the assumptions of
Section~\ref{sec:spread}.

We define the expected vulnerability of $\GCC{G}$, denoted by $\wPv$, as the
fraction of the nodes of $\GCC{G}$ that may become infected when the virus
attempts to infect a randomly chosen node of $\GCC{G}$. Let $\gcc{V}$ be the
fraction of the nodes of $G$ that belong to $\GCC{V}$. If the originator does
not belong to $\GIN{S}$ (which occurs with probability $1-\gin{S}/\gcc{G}$),
then $\wPv=1$; if it does belong to $\GIN{S}$ (with probability
$\gin{S}/\gcc{G}$), then $\wPv=\gcc{V}/\gcc{G}$ if and only if the virus first
infects a node of $\GCC{V}$, which occurs with probability $\gcc{V}/\gcc{G}$. We
then have
\begin{equation}
   \wPv = 1-\frac{\gin{S}}{\gcc{G}} + \frac{\gin{S}}{\gcc{G}}\left(\frac{\gcc{V}}{\gcc{G}}\right)^2.
   \label{eq:wPv}
\end{equation}

Henceforth in this section we concentrate on calculating $\gcc{V}$ for the case
in which $\GCC{V}$ does exist. Clearly, a node of $G$ belongs to $\GCC{V}$ only
if it does not belong to $\GOUT{S}$. Through the remainder of the section, let
$u$ be a degree-$a$ node of $G$ that does not belong to $\GOUT{S}$ and $v$ a
neighbor of $u$ in $G$. Given that $v$ has degree $b$, we define $\wH{a}{b}$ as
the probability that the edge $(v \to u)$ exists in $S$. Since $u$ does not
belong to $\GOUT{S}$, node $v$ must be such that it satisfies one of the
following conditions: either edge $(v \to u)$ does not exist in $S$, which
happens with probability $1-h(b,a)$, or $(v \to u)$ exists in $S$ but the number
of nodes that can reach $u$ through $v$ is small, which occurs with probability
$h(b,a)\qout{b}$. We can then express $\wH{a}{b}$ as the ratio of the
probability that the latter condition is satisfied to the probability that
either the former or the latter is. This leads to
\begin{equation}
   \wH{a}{b} = \frac{h(b,a)\qout{b}}{1 - h(b,a) + h(b,a)\qout{b}}.
   \label{eq:wh}
\end{equation}

Now let $\wP{a}{b}$ be the probability that $v$ has degree $b$ in $G$. Clearly,
$\wP{a}{b}$ is proportional to the joint probability that $v$ satisfies one of
the above conditions regarding the existence of edge $(v \to u)$ in $S$ and also
that a node's neighbor in $G$ has degree $b$. That is, $\wP{a}{b}$ is
proportional to
$\left(1-h(b,a)+h(b,a)\qout{b}\right)bP_G(b)/Z_G$. Using (\ref{eq:deout}), we
obtain
\begin{equation}
   \wP{a}{b} = \left(\frac{1-h(b,a)+h(b,a)\qout{b}}{\deout{a}}\right) \frac{bP_G(b)}{Z_G}.
   \label{eq:wp}
\end{equation}

Let $b$ be the degree of $v$ in $G$. Because $u$ does not belong to $\GOUT{S}$,
nodes $u$ and $v$ are neighbors in $V$ if and only if $v$ does not belong to
$\GOUT{S}$ either. If $(v \to u)$ is an edge of $S$, which occurs with
probability $\wH{a}{b}$, then $v$ is obviously not in $\GOUT{S}$, as it would
otherwise make $u$ belong to $\GOUT{S}$ along with it. On the other hand, if
$(v \to u)$ is not an edge of $S$ (with probability $1-\wH{a}{b}$), then $v$
does not belong to $\GOUT{S}$ if and only if the number of nodes that can reach
it in $S$ is small, which happens with probability $\qout{b}$. It follows that
the probability that $u$ and $v$ are neighbors in $V$ is given by
\begin{equation}
   \wH{a}{b} + (1 - \wH{a}{b})\qout{b}.
   \label{eq:aux30}
\end{equation}
When $u$ and $v$ are indeed neighbors in $V$, we define $\qgcc{b}$ as the
probability that $u$ has a small reach in $V$ through $v$. We say that $v$ is a
dead end with respect to $u$ in $V$ if either $v$ is not a neighbor of $u$ in
$V$, which occurs with probability
$1-\left[\wH{a}{b} + (1 - \wH{a}{b})\qout{b}\right]$, or it is but the reach of
$u$ through $v$ in $V$ is small, which occurs with probability
$\left[\wH{a}{b} + (1 - \wH{a}{b})\qout{b}\right]\qgcc{b}$. Thus, the
probability that $v$ is a dead end with respect to $u$ in $V$ is
\begin{eqnarray}
   \lefteqn{1-\left[\wH{a}{b} + (1 - \wH{a}{b})\qout{b}\right] +
   \left[\wH{a}{b} + (1 - \wH{a}{b})\qout{b}\right]\qgcc{b}}
   \hspace{1.25in}\nonumber\\
   && = \wH{a}{b}\qgcc{b} + (1-\wH{a}{b})(1 - \qout{b} + \qout{b}\qgcc{b}),
   \label{eq:wgccV}
\end{eqnarray}
so the probability that a neighbor of $u$ is a dead end with respect to $u$ in
$V$, which we denote by $\degcc{a}$, is clearly
\begin{equation}
   \degcc{a} = \sum_{b=1}^{n-1} \left[\wH{a}{b}\qgcc{b} + (1-\wH{a}{b})(1 - \qout{b} + \qout{b}\qgcc{b})\right]\wP{a}{b}.
   \label{eq:degccV}
\end{equation}

In order to calculate $\qgcc{b}$, notice that the reach of $u$ through $v$ in
$V$ is small if and only if all other $b-1$ neighbors of $v$ in $G$ are
themselves dead ends with respect to $v$ in $V$. Then, assuming that the degrees
of a node's neighbors in $G$ remain independent from one another even under the
condition that the node does not belong to $\GOUT{S}$, we have
\begin{equation}
   \qgcc{b} = (\degcc{b})^{b-1}.
   \label{eq:qgccV}
\end{equation}
Putting (\ref{eq:degccV}) and (\ref{eq:qgccV}) together leads to an equation
where $\degcc{a}$ is a function of all the other $\degcc{}$'s, which can then be
solved numerically for $0 \leq a \leq n-1$.

We are, finally, in position to calculate the value of $\gcc{V}$. Let $u$ be a
randomly chosen node of $G$ having degree $a$. In order to belong to $\GCC{V}$,
node $u$  must not belong to $\GOUT{S}$, which occurs with probability
$\left(\deout{a}\right)^a$. Furthermore, $u$ belongs to $\GCC{V}$ only if at
least one of its neighbors is not a dead end with respect to it in $V$, which
occurs with probability $1-(\degcc{a})^a$. It then follows that
\begin{equation}
   \gcc{V} = \sum_{a=0}^{n-1} (\deout{a})^a \left[1 - (\degcc{a})^a\right] P_G(a).
   \label{eq:gccV}
\end{equation}

\section{The heuristic function} \label{sec:heur}

The efficiency of heuristic flooding as a means of immunizing a network depends
heavily on the choice of the heuristic function $h(a,b)$. Before introducing our
heuristic function, we elaborate on the properties of subgraph $S$ that we may
expect to lead to good results for $\wPs$ and $\wPv$.

First of all, it is clear that $S$ must be above the phase transition that gives
rise to $\GSCC{S}$, thereby guaranteeing that $\GSCC{S}$, $\GIN{S}$, and
$\GOUT{S}$ almost surely exist. When this is the case, the nodes of $\GIN{S}$
are the most suitable ones for being the originator, as they can immunize a
non-negligible number of nodes. But since we cannot assume any prior information
on the originator, $\GIN{S}$ should contain as many nodes as possible in order
to make the probability that the originator is chosen from outside it as small
as possible. With regard to $\GOUT{S}$, we know that it contains the nodes that
receive the vaccine when the originator belongs to $\GIN{S}$. In order to
prevent an excessive number of nodes from receiving the vaccine, the size of
$\GOUT{S}$ should be kept to modest values. Putting these two observations
together, we ideally want $\GIN{S}$ to span all the nodes of the network,
$\GSCC{S}$ to contain only the nodes that can more efficiently block the
spreading of an infection, and $\GOUT{S}$ to be the same as $\GSCC{S}$.

Since we know that immunizing the nodes with the highest degrees is an efficient
way to prevent epidemics in scale-free networks
\cite{albert2000,cohen2001,satorras2003}, we introduce, in this section, a
heuristic function that stimulates the transmission of the vaccine to
high-degree nodes. Introducing a parameter $\alpha \geq 0$, and considering a
degree-$a$ node $u$ that has the vaccine and a degree-$b$ neighbor $v$ of $u$,
our heuristic function $h(a,b)$, which  gives the probability that $u$ sends the
vaccine to $v$, is defined as follows:
\begin{itemize}
\item If $b=1$, that is, $v$ has no neighbor besides $u$, then $h(a,b)=0$ and
$u$ deterministically decides not to send the vaccine to $v$. In this case,
since $u$ is already immune, should $v$ become infected it can transmit the
virus to no other node, so we choose not to give $v$ the vaccine.
\item If $a \leq 2 \leq b$, that is, $u$ has degree at most $2$ and $v$ has
degree at least $2$, then $h(a,b)=1$ and $u$ deterministically decides to send
the vaccine to $v$. This is meant to force some low-degree nodes to forward the
vaccine, thereby precluding a premature conclusion of heuristic flooding and, as
a consequence, leading to a larger $\GIN{S}$.
\item For all the other positive values of $a$ and $b$, we let 
\begin{equation}
   h(a,b)=\tanh\left(\frac{b-1}{\left(a-2\right)^\alpha}\right).
   \label{eq:h}
\end{equation}
\end{itemize}

\begin{figure*}[!t]
   \centering
   \begin{tabular}{c}
   \includegraphics[height=\heurHeight]{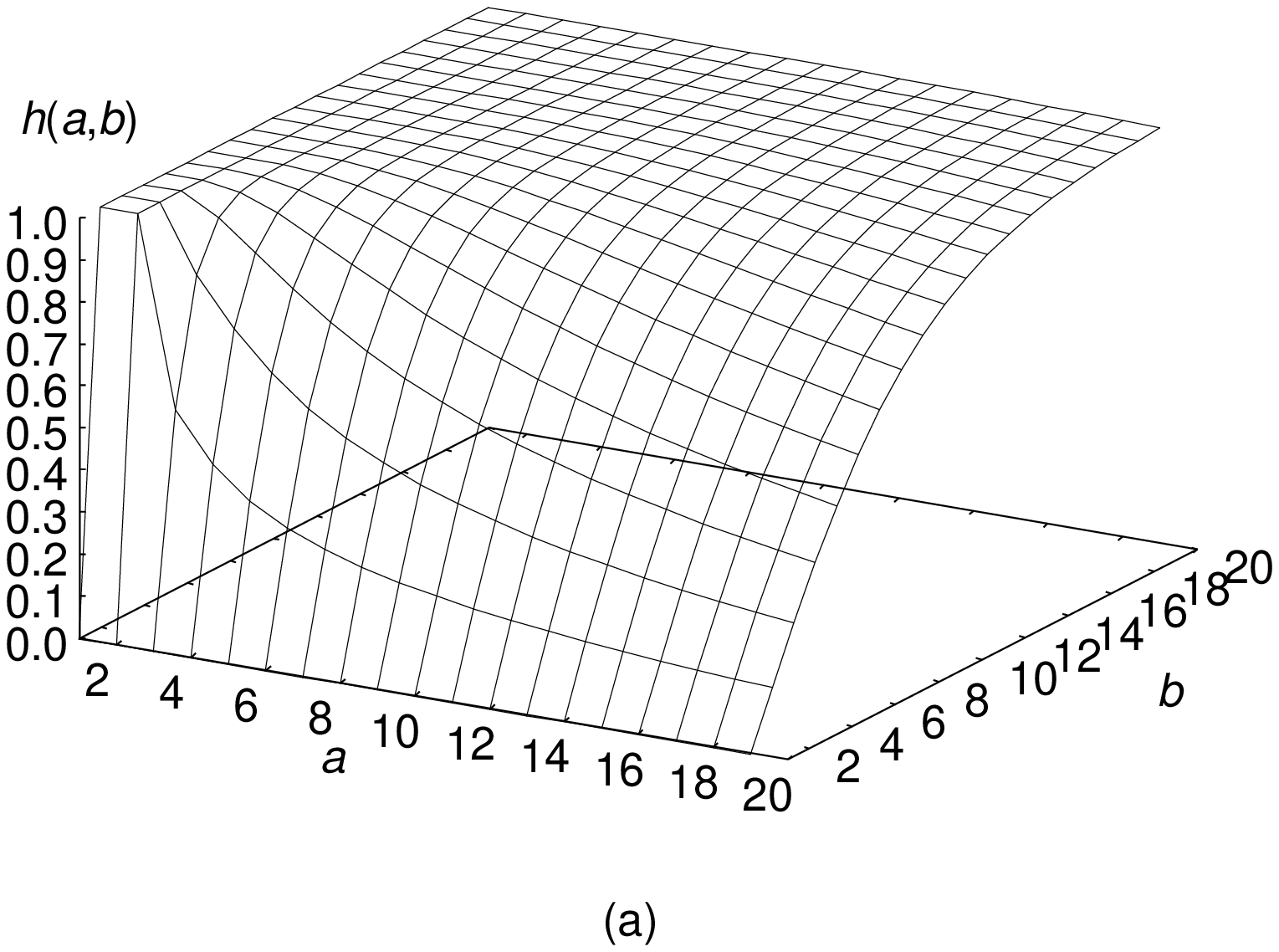}\\
   \\
   \includegraphics[height=\heurHeight]{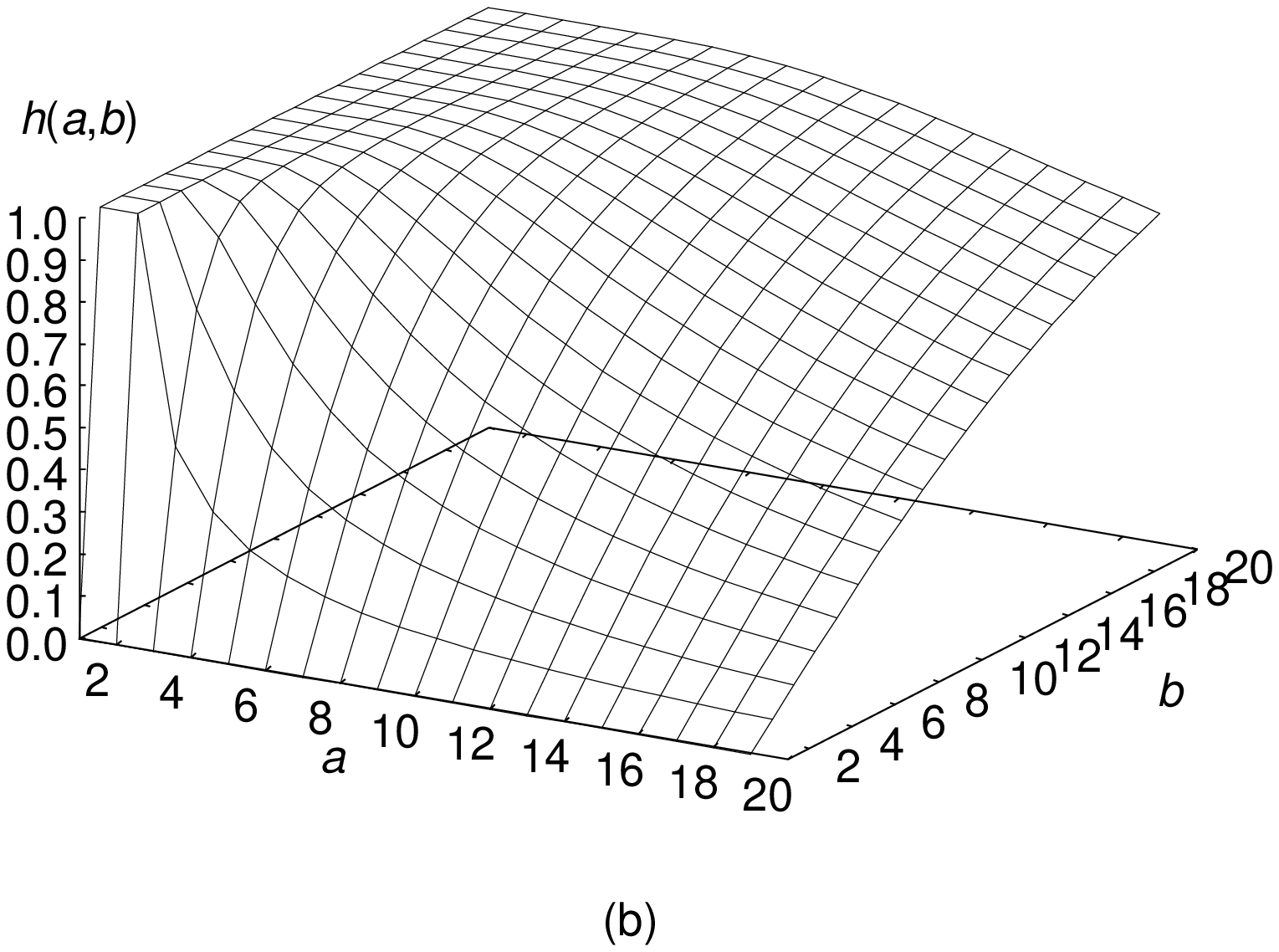}
   \end{tabular}
   \caption{Plots of the heuristic function given by (\ref{eq:h}) for
$\alpha=0.7$ (a) and $\alpha=1.0$ (b).}
   \label{fig:h}
\end{figure*}

Figure~\ref{fig:h} shows two plots illustrating this heuristic function for
$\alpha=0.7$ (part~(a)) and $\alpha=1.0$ (part~(b)). Clearly, for fixed $a>2$,
$h(a,b)$ increases with $b$, so the vaccine is more likely to be transmitted to
high-degree nodes. For fixed $b>1$, $h(a,b)$ decreases with $a$, thus reflecting
the intuition that, when $u$ is a high-degree node, sending the vaccine to $v$
may be unnecessary even if $v$ is a high-degree node (there are probably other
paths through which the vaccine can be transmitted from $u$ to $v$).

\section{Simulation results} \label{sec:sim}

We have conducted extensive simulations on random graphs with node degrees
distributed according to a power law. Generating such a graph is achieved in two
phases \cite{newman2001}. Let $u_1, u_2, \ldots, u_n$ be the nodes of the random
graph we want to generate. In the first phase, for $i=1,\ldots,n$ we sample the
degree $d_i$ of each $u_i$ from the power-law distribution, obtaining the
so-called degree sequence of the graph. If $\sum_{i=1}^n d_ i$ turns out to be
odd, then we discard the entire degree sequence and sample a new one, repeating
the process until the sum of the degrees comes out even. In the second phase, we
consider an imaginary urn having $\sum_{i=1}^n d_i$ labeled balls, the labels of
$d_i$ of them being $u_i$. We then successively remove pairs of balls from the
urn until it has no more balls. For each pair we remove---say, of labels $u_i$
and $u_j$---we add edge $(u_i,u_j)$ to the graph. This method can produce graphs
having multiple edges or self-loops, but it has the advantage of generating
graphs whose degrees remain independent even after the edges are added, which is
a core assumption of our analysis.

We carried out our simulations for $n=10000$ and $2 \leq \tau \leq 3$. For each
value of $\tau$, we generated $500$ $G$ instances. For each $G$ instance, we
used the heuristic $h(a,b)$ to both sample $1000$ instances of the subgraph $S$
and, in an independent way, conduct $1000$ vaccine disseminations by heuristic
flooding from an originator selected randomly among the nodes of the largest
connected component of the $G$ instance. For each $S$ instance, we selected the
largest strongly connected component and calculated the sizes of the
corresponding in-component (counting the nodes that can reach the strongly
connected component) and out-component (counting the nodes that can be reached
from the strongly connected component). We then obtained the expected sizes of
$\GIN{S}$ and $\GOUT{S}$ by averaging these quantities over the $500000$
samples. For each vaccine dissemination, we calculated the fraction of nodes
that receive the vaccine and the fraction of nodes to which an infection may
spread when an attempt at infecting a randomly chosen node inside the largest
connected component of $G$ takes place. We then obtained $\wPs$ and $\wPv$ by
averaging these quantities over the $500000$ samples.

\begin{figure*}[!t]
   \centering
   \begin{tabular}{rr}
   \includegraphics[height=\graphHeight]{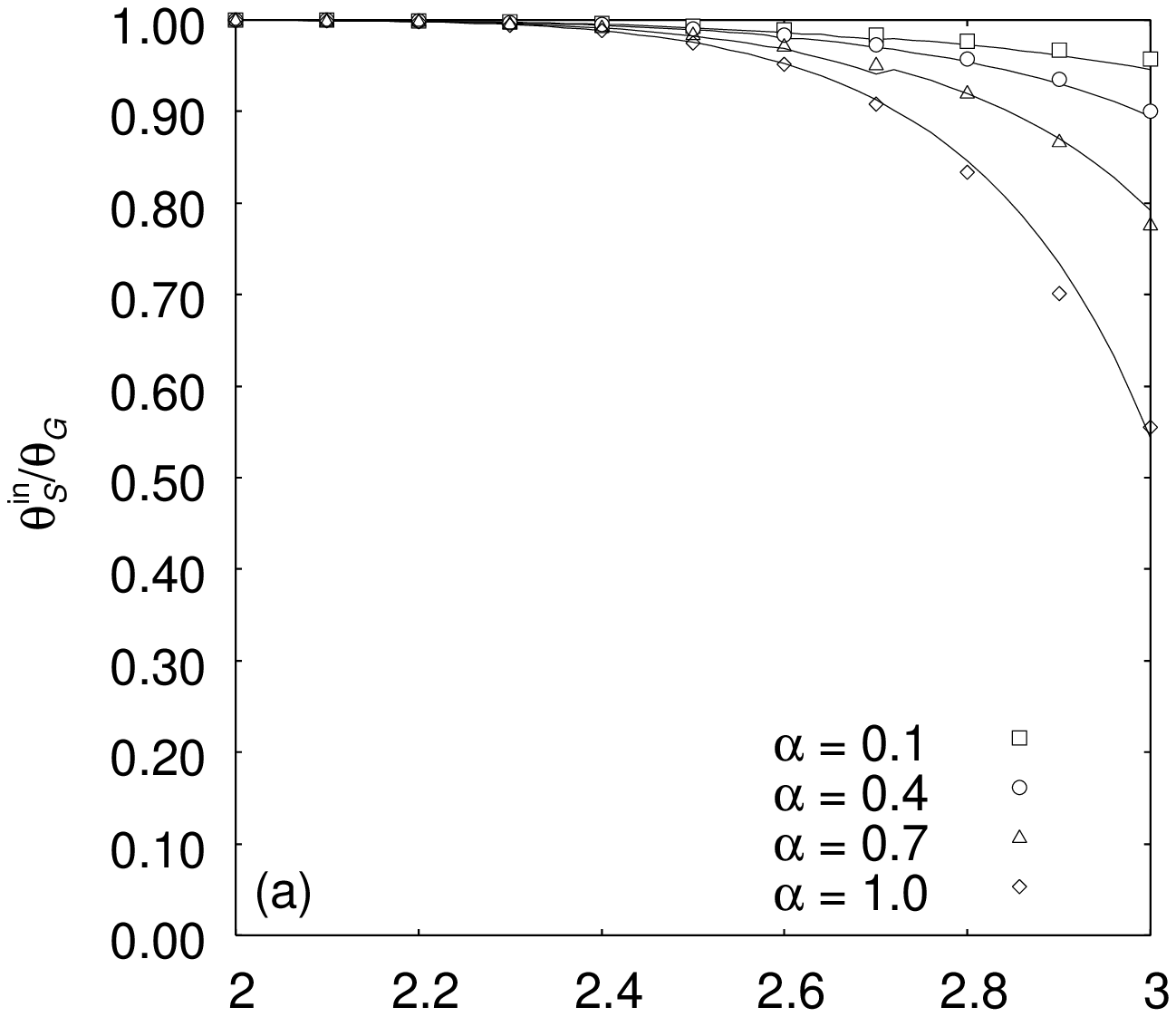} &
   \includegraphics[height=\graphHeight]{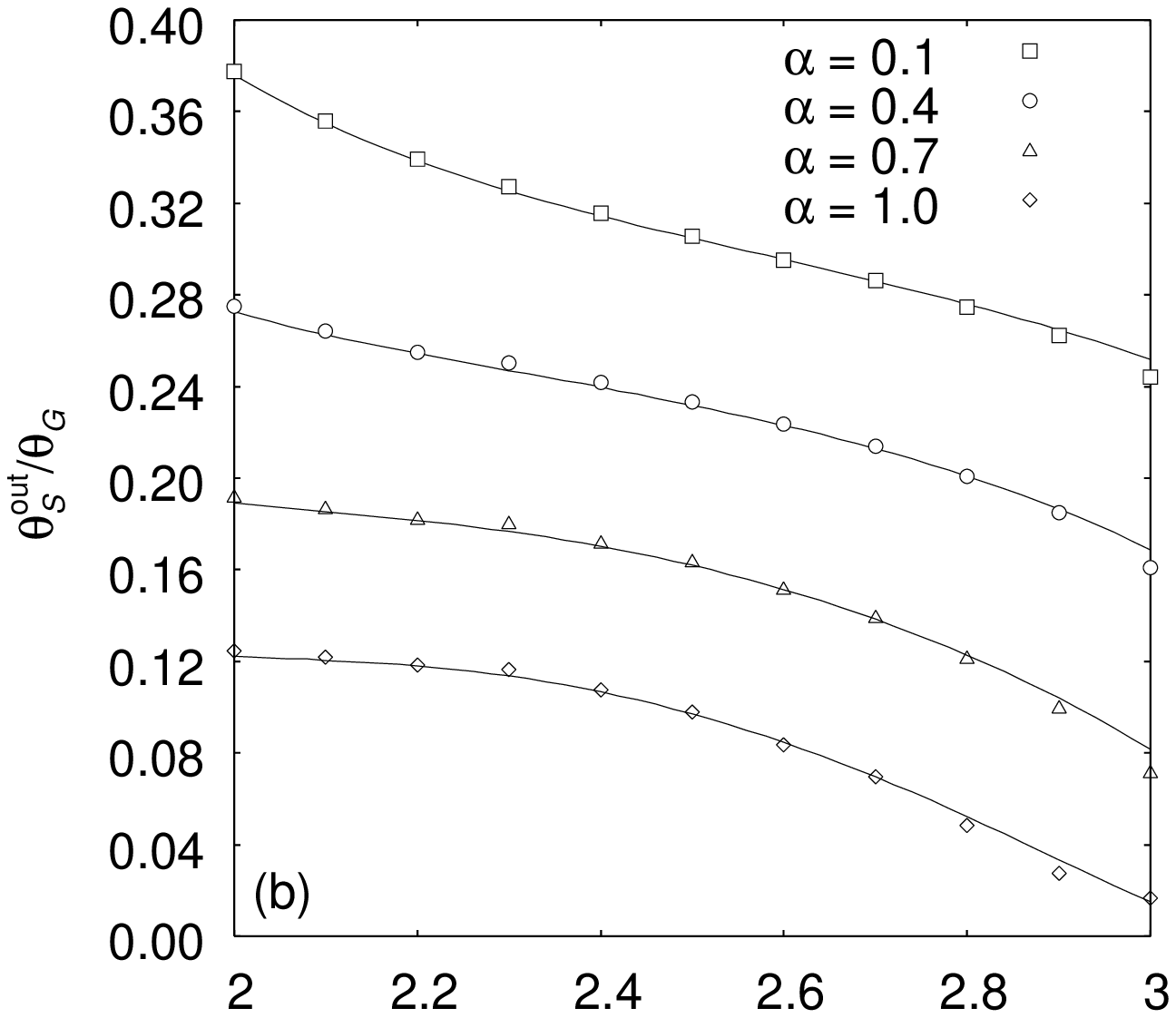} \\
   \includegraphics[height=\graphHeight]{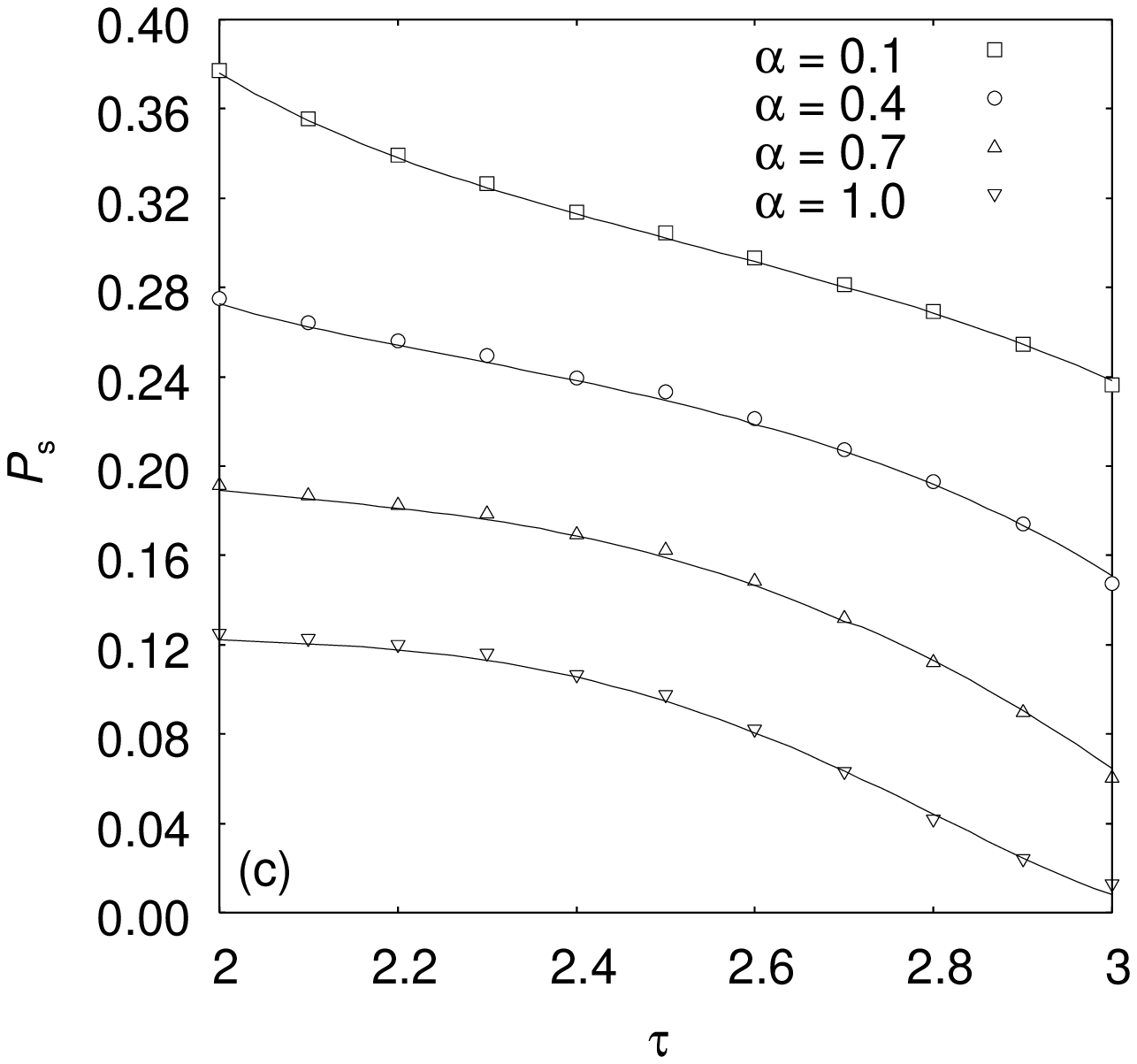} &
   \includegraphics[height=\graphHeight]{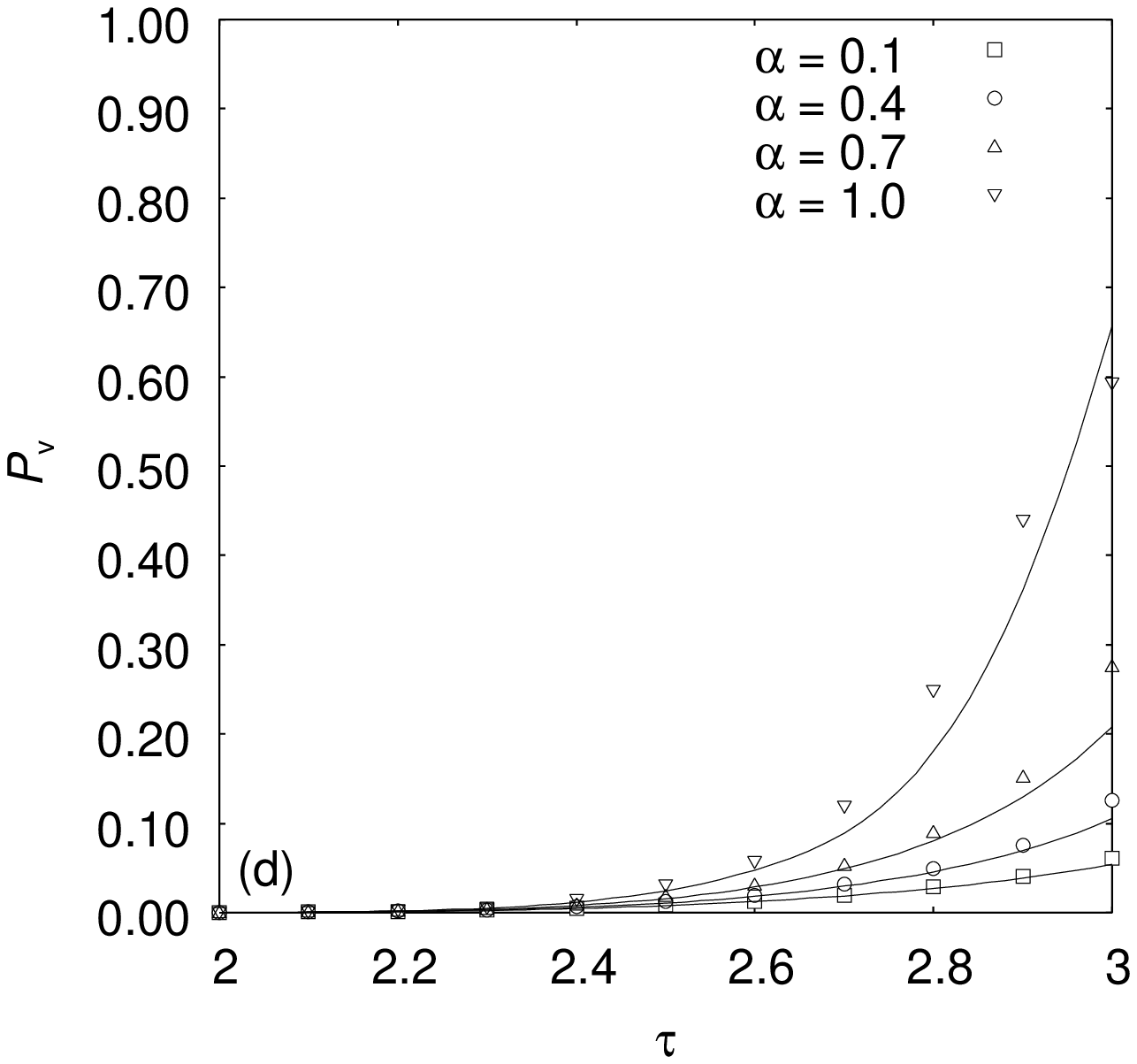}
   \end{tabular}
   \caption{Simulation results of vaccine dissemination by heuristic flooding.
Solid lines give the analytic predictions.}
   \label{fig:sim}
\end{figure*}

Simulation results are shown in Figure~\ref{fig:sim} for
$\alpha=0.1,0.4,0.7,1.0$. We note, in general, a satisfactory agreement between
analytic and simulation results, with the exception of part~(d), in which case
the deviation may be attributed to the approximations made during the derivation
of $\gcc{V}$ in Section~\ref{sec:vulnerability} to yield (\ref{eq:gccV}).

When $\tau \leq 2.5$, the plots for $\gin{S}/\gcc{G}$ and $\gout{S}/\gcc{G}$
(Figure~\ref{fig:sim}(a,b)) reveal that the heuristic function introduced in
Section~\ref{sec:heur} results in a $\GIN{S}$ that spans almost all the nodes of
$\GCC{G}$, while the size of $\GOUT{S}$ keeps to a relatively modest fraction of
$\GCC{G}$. For example, for $\tau \leq 2.5$ and $\alpha=1.0$, the relative size
of $\GIN{S}$ is always above $0.97$ and the relative size of $\GOUT{S}$ is
always below $0.13$. For $\tau > 2.5$, the relative size of $\GIN{S}$ decreases
with $\tau$, thus evidencing that heuristic flooding has more difficulty
disseminating the vaccine when the graph is sparser.

Owing to $\wPs$ being given by $(\gin{S}/\gcc{G})(\gout{S}/\gcc{G})$
(cf.\ (\ref{eq:wPi})), and to $\gin{S}/\gcc{G}$ being relatively close to $1$
(Figure~\ref{fig:sim}(a)), the plots for $\wPs$ (Figure~\ref{fig:sim}(c)) are of
course similar to the plots for $\gout{S}/\gcc{G}$ (Figure~\ref{fig:sim}(b)).
Furthermore, given a value of $\alpha$, $\wPs$ decreases with $\tau$, which
means that heuristic flooding spreads through a smaller number of nodes when the
graph is sparser, as, in this case, there are less paths conducting to the
high-degree nodes.

As for $\wPv$ (Figure~\ref{fig:sim}(d)), we note that, for $\tau \leq 2.5$,
$\wPv$ is nearly zero. This result is a natural consequence both of the guiding
principle of the heuristic introduced in Section~\ref{sec:heur}, which ascribes
more probability for transmitting the vaccine to nodes having higher degrees,
and of the result for $\gin{S}/\gcc{G}$ (Figure~\ref{fig:sim}(a)), which
indicates that $\GIN{S}$ spans almost all the nodes of $\GCC{G}$. As $\tau$ is
increased to values greater than $2.5$, $\wPv$ moves farther away from zero,
since the size of $\GIN{S}$ decreases and, therefore, the probability that
heuristic flooding distributes the vaccine to only a small number of nodes
increases. Regarding the value of $\alpha$, we note a clear trade-off between
$\wPs$ and $\wPv$. If we were to adjust $\alpha$ in such a way as to decrease
$\wPs$, we would have an increase in $\wPv$, which shows that the number of
immunized nodes has a direct impact on the resulting vulnerability of the
network.

\section{Conclusion} \label{sec:conc}

We have considered in this paper the problem of immunizing a scale-free network
against a virus or worm. We introduced a new immunization strategy, one that we
believe reflects more accurately what happens in real scenarios. In our
strategy, we assume that the vaccine enters the network at exactly one node, in
general the site of the vaccine's development or the site in charge of its
distribution, for example. This node begins the dissemination of the vaccine by
heuristic flooding, aiming at immunizing the nodes that have the highest
degrees. With this purpose in mind, we introduced a heuristic function that
gives more probability to forwarding the vaccine toward nodes with higher
degrees.

We obtained analytical and simulation results on random graphs having node
degrees distributed according to a power law. Our mathematical analysis has
innovative aspects that we expect may shed some light on obtaining analytical
results for similar distributed algorithms. Also, we hope our analysis can
contribute to the development of new heuristic functions for vaccine
dissemination. With regard to our simulation results, they show satisfactory
agreement with our mathematical analysis and highlight the expected trade-off
between the number of nodes that receive the vaccine and the vulnerability of
the network to future infections. Especially for power laws with relatively
small value for the parameter $\tau$, our heuristic function achieves very good
results, making the network practically invulnerable to an epidemic while
requiring the immunization of only roughly $10\%$ of the nodes.

We note, finally, that one possible direction in which this paper's research may
be extended, in addition to the search for other heuristic functions, is that of
allowing for multiple concurrent initiators. While algorithmically (i.e., from
the perspective of flooding the network) such an extension is trivial, extending
the analysis of Section~\ref{sec:math} is expected to be a significantly more
complex endeavor.

\subsection*{Acknowledgments}

The authors acknowledge partial support from CNPq, CAPES, and a FAPERJ BBP
grant.

\bibliography{imm}

\begin{thebibliography}{10}

\bibitem{albert2002}
R.~Albert and A.-L. Barab{\'a}si.
\newblock Statistical mechanics of complex networks.
\newblock {\em Reviews of Modern Physics}, 74:47--97, 2002.

\bibitem{albert2000}
R.~Albert, H.~Jeong, and A.-L. Barab{\'a}si.
\newblock Error and attack tolerance of complex networks.
\newblock {\em Nature}, 406:378--382, 2000.

\bibitem{barabasi1999}
A.-L. Barab{\'a}si and R.~Albert.
\newblock Emergence of scaling in random networks.
\newblock {\em Science}, 286:509--512, 1999.

\bibitem{bollobas2002}
B.~Bollobás.
\newblock {\em Random Graphs}.
\newblock Cambridge University Press, Cambridge, UK, 2 edition, 2001.

\bibitem{cohen2000}
R.~Cohen, K.~Erez, D.~{ben-Avraham}, and S.~Havlin.
\newblock Resilience of the {I}nternet to random breakdowns.
\newblock {\em Physical Review Letters}, 85:4626--4628, 2000.

\bibitem{cohen2001}
R.~Cohen, K.~Erez, D.~{ben-Avraham}, and S.~Havlin.
\newblock Breakdown of the {Internet} under intentional attack.
\newblock {\em Physical Review Letters}, 86:3682--3685, 2001.

\bibitem{cohen2003}
R.~Cohen, S.~Havlin, and D.~{ben-Avraham}.
\newblock Efficient immunization strategies for computer networks and
  populations.
\newblock {\em Physical Review Letters}, 91:247901, 2003.

\bibitem{erdos1959}
P.~Erd{\H o}s and A.~R{\'e}nyi.
\newblock On random graphs.
\newblock {\em Publicationes Mathematicae}, 6:290--297, 1959.

\bibitem{faloutsos1999}
M.~Faloutsos, P.~Faloutsos, and C.~Faloutsos.
\newblock On power-law relationships of the {I}nternet topology.
\newblock In {\em Proceedings of the Conference on Applications, Technologies,
  Architectures, and Protocols for Computer Communication}, pages 251--262,
  1999.

\bibitem{molloy1995}
M.~Molloy and B.~Reed.
\newblock A critical point for random graphs with a given degree sequence.
\newblock {\em Random Structures and Algorithms}, 6:161--180, 1995.

\bibitem{molloy1998}
M.~Molloy and B.~Reed.
\newblock The size of the largest component of a random graph on a fixed degree
  sequence.
\newblock {\em Combinatorics, Probability and Computing}, 7:295--306, 1998.

\bibitem{newman2003b}
M.~E.~J. Newman.
\newblock The structure and function of complex networks.
\newblock {\em SIAM Review}, 45:167--256, 2003.

\bibitem{newman2001}
M.~E.~J. Newman, S.~H. Strogatz, and D.~J. Watts.
\newblock Random graphs with arbitrary degree distributions and their
  applications.
\newblock {\em Physical Review E}, 64:026118, 2001.

\bibitem{satorras2001}
R.~Pastor-Satorras and A.~Vespignani.
\newblock Epidemic spreading in scale-free networks.
\newblock {\em Physical Review Letters}, 86:3200--3203, 2001.

\bibitem{satorras2003}
R.~Pastor-Satorras and A.~Vespignani.
\newblock Epidemics and immunization in scale-free networks.
\newblock In S.~Bornholdt and H.~G. Schuster, editors, {\em Handbook of Graphs
  and Networks: From the Genome to the Internet}, pages 111--130. {Wiley-VCH},
  Weinheim, Germany, 2003.

\bibitem{stauffer2004}
A.~O. Stauffer and V.~C. Barbosa.
\newblock Probabilistic heuristics for disseminating information in networks,
  2004.
\newblock http://arxiv.org/abs/cs.NI/0409001.

\end{thebibliography}
\bibliographystyle{plain}

\end{document}